\documentclass[aip,pof,reprint,onecolumn]{revtex4-1}
\usepackage{amsmath}
\usepackage{graphicx}
\usepackage{color}
\usepackage{hyperref}
\usepackage{psfrag}
\usepackage{stmaryrd}
\usepackage{amssymb}
\usepackage{wasysym}
\usepackage[latin1]{inputenc}

\def \degre {$^\mathrm{o}$}

\usepackage{geometry}
\newcommand{\cor}[1]{\textcolor{black}{#1}}

\begin{document}

\title{Direct and inverse energy cascades in a forced rotating turbulence experiment}

\author{Antoine Campagne}
\affiliation{Laboratoire FAST,
CNRS, Universit\'e Paris-Sud, 91405 Orsay, France}
\author{Basile Gallet}
\affiliation{Laboratoire FAST, CNRS,
Universit\'e Paris-Sud, 91405 Orsay, France}
\affiliation{Laboratoire SPHYNX, Service de Physique de l'\'Etat Condens\'e, DSM, CEA Saclay, CNRS, 91191 Gif-sur-Yvette, France}
\author{Fr\'{e}d\'{e}ric Moisy}
\affiliation{Laboratoire FAST, CNRS,
Universit\'e Paris-Sud, 91405 Orsay, France}
\author{Pierre-Philippe Cortet}
\affiliation{Laboratoire FAST,
CNRS, Universit\'e Paris-Sud, 91405 Orsay, France}

\date{\today}

\begin{abstract}

We present experimental evidence for a double cascade of kinetic
energy in a statistically stationary rotating turbulence
experiment. Turbulence is generated by a set of vertical flaps
which continuously injects velocity fluctuations towards the
center of a rotating water tank. The energy transfers are
evaluated from two-point third-order three-component velocity
structure functions, which we measure using stereoscopic particle
image velocimetry in the rotating frame. Without global rotation,
the energy is transferred from large to small scales, as in
classical three-dimensional turbulence. For nonzero rotation
rates, the horizontal kinetic energy presents a double cascade: a
direct cascade at small horizontal scales and an inverse cascade
at large horizontal scales. By contrast, the vertical kinetic
energy is always transferred from large to small horizontal
scales, a behavior reminiscent of the dynamics of a passive scalar
in two-dimensional turbulence. At the largest rotation rate the
flow is nearly two-dimensional, and a pure inverse energy cascade
is found for the horizontal energy. To describe the scale-by-scale
energy budget, we consider a generalization of the
Kármán-Howarth-Monin equation to inhomogeneous turbulent flows, in
which the energy input is explicitly described as the advection of
turbulent energy from the flaps through the surface of the control
volume where the measurements are performed.

\end{abstract}

\maketitle

\section{Introduction}

Global rotation is a key ingredient of many geophysical and
astrophysical flows. Through the action of the Coriolis force,
rotating turbulence tends to approach two-dimensionality, i.e.
invariance along the rotation axis (hereafter denoted as the
vertical axis by
convention).\cite{CambonBook,DavidsonBook2013,Godeferd2014}
Energetic 2D and 3D flow features therefore coexist in rotating
turbulence, and the question of the direction of the energy
cascade between spatial scales naturally arises: In 3D, energy is
transferred from large to small
scales\cite{CambonBook,DavidsonBook2013,FrischBook} whereas it is
transferred from small to large scales in 2D, as first proposed by
Kraichnan.\cite{Kraichnan1967,Lindborg1999,Tabeling2002} In
rotating turbulence, energy transfers depend on the Rossby number
$Ro$, which compares the rotation period $\Omega^{-1}$ to the
turbulent turnover time. In the limit of small $Ro$, the fluid
motions evolving on a time scale much slower than the rotation
period $\Omega^{-1}$ are 2D3C (two-dimensional, three-component),
a result known as the Taylor-Proudman theorem, while the faster
motions of frequency up to $2 \Omega$ are in the form of 3D
inertial waves.\cite{GreenspanBook} In this limit, 3D energy
transfers occur through resonant and quasi-resonant triadic
interactions of inertial
waves,\cite{Cambon1989,Waleffe1992,Waleffe1993,Bordes2012,Smith1999}
which drive energy in a {\it direct} cascade, with a net transfer
towards slow, small-scale, nearly 2D
modes.\cite{Waleffe1993,Galtier2003,Cambon2004} Exactly resonant
triads cannot however drive energy from 3D modes to the exactly 2D
mode. In the limit of vanishing Rossby number, only those exact
resonances are efficient, so the 2D3C mode is
autonomous:\cite{Bourouiba2008} It follows a purely 2D dynamics
unaffected by rotation, with an inverse cascade of horizontal
energy and a passive-scalar mixing of the vertical
velocity.\cite{FrischBook,Tabeling2002} This decoupling implies
that, if energy is supplied to the 3D modes only, the 2D mode
should not be excited and no inverse cascade should be observed.

In contrast with this asymptotic limit, most experiments and
numerical simulations correspond to moderate Rossby numbers. They
exhibit the emergence of large-scale columnar structures, which
suggests a net transfer from the 3D ``wave'' modes to the 2D3C
``vortex''
mode.\cite{Hopfinger1982,Bartello1994,Bourouiba2007,Mininni2009,Mininni2010,Pouquet2010,Moisy2011,Bourouiba2012,Pouquet2013}
\cor{For such moderate Rossby numbers, near-resonant triadic
interactions, which are increasingly important as $Ro$ is
increased, allow for non-vanishing energy transfers between 3D and
2D
modes,\cite{Waleffe1993,Chen2005,Smith2005,Bourouiba2007,Bourouiba2012}
thus providing a mechanism for the emergence of inverse energy
transfers: once energy is transferred to the 2D vortex mode, local
2D interactions are expected to build an upscale energy cascade.
Even for a purely 3D forcing, the vortex mode grows as a result of
near-resonant triads involving one 2D mode and two
large-vertical-scale and small-horizontal-scale 3D
modes:\cite{Bourouiba2012} this vortex mode then triggers inverse
energy transfers between purely 2D modes.} This intermediate
Rossby number regime is of first practical interest: the Rossby
number of most laboratory experiments and
geophysical/astrophysical flows is indeed of the order of $10^{-1}
- 10^{-2}$. In these situations, a natural question is to what
extent direct and inverse cascades may coexist, and what sets
their relative amplitudes as the Rossby number is varied.

Inverse energy cascade in rotating turbulence has been mostly
investigated numerically, in the simplified configuration of a
body force acting at an intermediate wave number $k_f$ in a
periodic
box.\cite{Hossain1994,Smith1996,Yeung1998,Smith1999,Chen2005,Mininni2009,Sen2012,Pouquet2013,Deusebio2014}
In this setup the inverse cascade is manifested through a growth
of the energy spectrum, and hence an inverse spectral transfer, at
wave numbers $k_\perp<k_f$ (with $k_\perp$ the wave number
component normal to the rotation axis). As for 2D turbulence, the
kinetic energy increases during this transient regime, until
energetic domain size structures are formed\cite{Gallet2013} or
additional large-scale dissipation comes into play. Although much
weaker, an inverse transfer of energy is also found in numerical
simulations of decaying rotating
turbulence.\cite{Bourouiba2007,Thiele2009,Teitelbaum2011} Overall,
these simulations indicate that, in addition to the Rossby number,
the nature of the forcing, in particular its {\it dimensionality}
(2D vs. 3D), {\it componentality} (2C vs. 3C) and helicity
content, play key roles for the existence and intensity of the
inverse cascade.\cite{Smith2005,Sen2012} In addition, since
shallow domains resemble 2D systems, which enhances the inverse
cascade, another key parameter in this problem is the vertical
confinement: the critical Rossby number under which the inverse
cascade appears increases as the ratio of the box height to the
forcing scale gets smaller.\cite{Smith1996,Smith1999,Deusebio2014}

\cor{Although these numerical simulations have provided valuable
insight about the conditions under which an inverse cascade takes
place in rotating turbulence, the most common assumptions of
homogeneity and narrow-band spectral forcing are of limited
practical interest. More general forcing functions are considered
in the simulations of Bourouiba \textit{et
al.},\cite{Bourouiba2012} with energy input either in a large
range of vertical scales and a single horizontal scale, or
vice-versa. In most flows encountered in the laboratory and in
geophysical/astrophysical contexts, energy injection in a given
control volume is broadband and results from the spatial gradients
of turbulent energy.} As a consequence, the well-separated inverse
and direct cascades obtained in numerical simulations with a
separating wave number fixed at the forcing wave number $k_f$ are
not relevant to describe real flows with boundary forcing.
Furthermore, flows of geophysical relevance can often be
considered to be in statistically steady state. Such stationary
states are easily achieved in laboratory experiments, whereas they
generally correspond to prohibitively long integration times for
numerical simulation. This provides another justification for
considering the problem of the energy cascade directions of
rotating turbulence experimentally.

We therefore built an experiment aimed at studying such stationary
rotating turbulence. Designing a rotating turbulence experiment
which unambiguously exhibits an inverse cascade is however
difficult for several reasons. First, in a statistically steady
turbulence experiment, an inverse cascade can be identified only
from measurements of energy transfers, i.e. from third-order
velocity correlations. These measurements require very large data
sets from advanced image-based diagnostic such as stereoscopic
particle image velocimetry.\cite{Lamriben2011} Second, it is
possible to separate the scale-by-scale energy fluxes from the
spatial transport of energy only under the assumption of weak
inhomogeneity of the flow, which is difficult to satisfy with
boundary forced experiments.

Because of these difficulties, experimental evidence of inverse
cascade in rotating turbulence is scarce. Indirect evidence was
first provided by Baroud~\textit{et al.}\cite{Baroud2002} in
forced turbulence and later by Morize~\textit{et
al.}\cite{Morize2005} in decaying turbulence. In both
experiments it is reflected in a change of sign of the third-order
moment of the longitudinal velocity increments in the plane normal
to the rotation axis, $S_3(r_\perp) = \langle (\delta u_L)^3
\rangle$ (where $u_L$ is the velocity increment projected along
the horizontal separation ${\bf r}_\perp$). Simple relations
between $S_3(r_\perp)$ and the energy flux exist only either in
the 3D3C isotropic case or in the 2D2C isotropic case, but not in
the general axisymmetric case, so the change of sign of $S_3$
cannot be unambiguously related  to inverse energy transfers
in these experiments. More recently, evidence of inverse energy
transfers has been reported by Yarom {\it et
al.},\cite{Yarom2013} from the transient evolution of the energy
spectrum in a forced rotating turbulence experiment. However,
because of the unstationnary and inhomogeneous nature of their
experiment, it is delicate to distinguish the {\it scale-by-scale}
energy transfers at a given spatial location from the {\it
spatial} energy transport from the turbulence production device to
the measurement area. In all these experiments the aspect ratio is
of order unity, so the 2D features of turbulence are essentially
due to rotation and not confinement. The extreme case of
rotating shallow water experiments is indeed known to produce a
purely 2D dynamics with an inverse energy cascade even at small
rotation rate (see, e.g., Afanasyev and
Craig\cite{Afanasyev2013}). The integral scales measurements of
van Bokhoven {\it et al.},\cite{Bokhoven2009} in which both the
fluid height and the rotation rate are varied, also confirm the
combined roles of these two parameters in the generation of
large-scale quasi-2D vortices.

In this paper we investigate the interplay between direct and
inverse energy cascades in a statistically stationary rotating
turbulence experiment from direct measurements of scale-by-scale
energy transfers. Turbulence is generated by a set of vertical
flaps which continuously inject velocity fluctuations towards the
center of a rotating water tank. The flaps are vertically
invariant, but instabilities in their vicinity induce 3D turbulent
fluctuations, so the forcing injects energy both in the 2D
and 3D modes. We compute the energy transfers from the
divergence of the two-point third-order velocity structure
functions extracted from stereoscopic particle image velocimetry
measurements in the rotating frame. We observe the emergence of a
double cascade of energy, direct at small scales and inverse at
large scales, the extension and magnitude of the inverse cascade
increasing with global rotation. This overall behavior of the
total kinetic energy is the superposition of different behaviors
for the horizontal and vertical velocities: for rapid global
rotation, the horizontal energy exhibits an inverse cascade,
whereas the vertical energy follows a direct cascade. The inverse
cascade of horizontal energy is found only at large scale for
moderate rotation rate, but gradually spreads down to the smallest
scales as the rotation rate is increased. These findings are
compatible with a 2D3C dynamics at large rotation rate, with the
horizontal velocity following a 2D dynamics and the vertical
velocity behaving as a passive scalar.

The energy transfers in homogeneous (but not necessarily
isotropic) turbulence can be described in the physical space using
the Kármán-Howarth-Monin (KHM)
equation.\cite{Monin1975,FrischBook,Lamriben2011} This approach
holds for homogeneous decaying turbulence and for stationary
turbulence forced by a homogeneous body force. However, it breaks
down in boundary-forced experiments, in which inhomogeneities
induce a transport of kinetic energy from the forcing region to
the region where measurements are performed. Extended versions of
the KHM equation including the effects of inhomogeneities have
been proposed and proved useful to describe the energy budget in
simple configurations, e.g. in wind-tunnel
experiments.\cite{Lindborg1999b,Hill2002,Antonia2006,Danaila2012a,Danaila2012b}
Here we make use of the inhomogeneous generalization of the KHM
equation proposed by Hill.\cite{Hill2002}  The measurement of the
different terms of this equation in the case of the largest
rotation rate, which is closer to the asymptotic 2D3C state,
allows us to clarify the effect of the inhomogeneous forcing in
this experiment.

\section{Experimental Setup}
\label{sec:setup}

\begin{figure}
\centerline{\includegraphics[width=13cm]{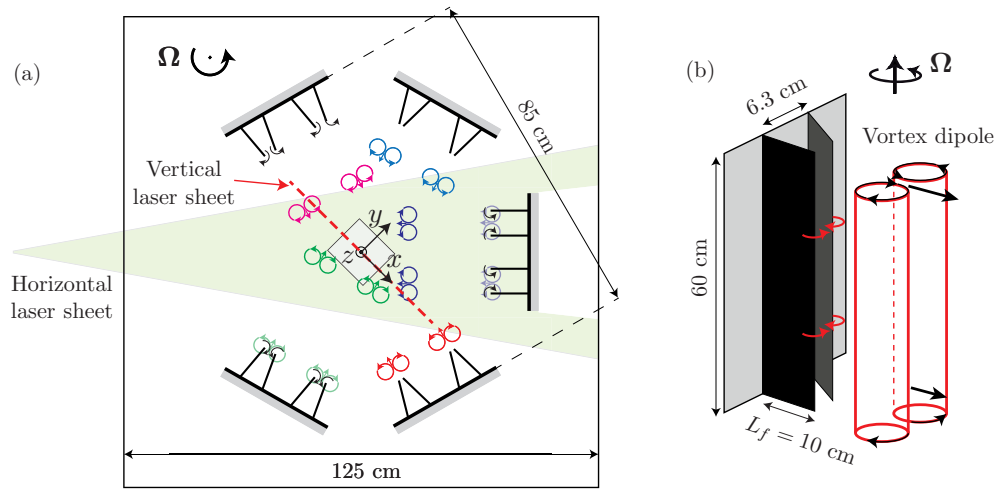}}
\caption{(Color online) (a) Schematic of the experiment: an arena
of 10 pairs of vertical flaps is placed in a parallelepipedic
water tank rotating at angular velocity $\Omega$. The rotation
vector ${\bf \Omega}$ is vertical and the system is viewed from
above. The rectangle at the center of the arena indicates the
horizontal region where 2D-3C velocity fields are measured by
stereoscopic particle image velocimetry. The drawing shows
idealized vortex dipoles emitted by the generators, before they
interact in the center of the arena. (b) Perspective view of a
pair of flaps.}\label{fig:arene}
\end{figure}

The experimental setup is similar to the one described in
Gallet~{\it et al.},\cite{Gallet2014} and only the features
specific to the present experiments are described in detail here.
The setup consists of a glass tank of $125\times125$~cm$^2$ square
base and $65$~cm height, filled with $50$~cm of water and mounted
on a precision rotating platform of 2~m diameter (see
Fig.~\ref{fig:arene}(a)). We have carried out experiments at five
rotation rates $\Omega$ in the range $0.21$ to $1.68$~rad~s$^{-1}$
($2$ to $16$~rpm), together with a reference experiment without
rotation ($\Omega=0$). The rotation rate is constant to better
than $10^{-3}$ relative fluctuations. In the central region of the
tank, we use a glass lid to avoid the paraboloidal deformation of
the free surface and to allow for visualization from above. This
lid is $43$~cm above the bottom of the tank.

A statistically stationary turbulent flow is produced by a set of
ten vortex dipole generators. They are arranged in 5 blocks of 2
generators located around a hexagonal arena of $85\pm 5$~cm width,
each of them being oriented towards the center of the arena
(Fig.~\ref{fig:arene}(a)). One side of the hexagon is left open to
illuminate the center of the arena with a horizontal laser sheet.
This forcing device was initially designed to generate turbulence
in stratified fluids, and is described in detail in
Refs.~\onlinecite{Billant2000,Augier2014}. Each generator consists
of a pair of vertical flaps, 60~cm high and $L_f=10$~cm long, each
flap rotating about one of its vertical edges
(Fig.~\ref{fig:arene}(b)). Thanks to DC motors and a system of
gears and cams, the pairs of flaps are driven in a periodic motion
of $8.5$~s duration and 9\degre~amplitude: the two flaps being
initially parallel, they first rotate with an angular velocity
$\omega_f = 0.092$~rad~s$^{-1}$ during $1.7$~s until their tips
almost touch each other. They remain motionless during $3.4$~s,
before reopening during 1.7~s until they reach the initial
parallel configuration again. They finally remain motionless
during the last 1.7~s of the cycle. The motions of the two
adjacent pairs of flaps of a given block are in phase, but an
arbitrary phase shift is set between the five blocks. The rotation
of the platform is set long before the start of this forcing
device, at least 1~h, in order for transient spin-up
recirculations to be damped. Once solid-body rotation is reached,
we start the generators, and a statistically stationary state is
reached in the center of the tank after a few minutes.

The Reynolds number based on the flap length $L_f$ and flap
angular velocity $\omega_f$ is $Re_f=\omega_f L_f^2/\nu = 920$.
The flow generated by the closing of the flaps consists of an
initially vertically-invariant vortex dipole
(Fig.~\ref{fig:arene}(a)) which quickly becomes unstable and
produces small-scale 3D turbulent fluctuations. This turbulent
burst self-advects towards the center of the arena because of the
persistent large-scale vortex dipole component. The Rossby number
based on the flap angular velocity is low,
$Ro_f=\omega_f/2\Omega\in [0.03, 0.22]$ (see Tab.~\ref{tab:Re}),
indicating that the flow generated by the flap motion is
influenced by rotation right from the generators (except for the
non-rotating experiment).  Turbulence in the center of the flow
can be also characterized locally by the turbulent Reynolds and
Rossby numbers based on the r.m.s. velocity $u_{\rm rms}$ and the
horizontal integral scale $L_\perp$ defined in
Eq.~(\ref{eq:Lperp}) (see next Section): $Re=u_{\rm rms}
L_\perp/\nu$ and $Ro = u_{\rm rms} /2\Omega L_\perp$ (values are
given in Tab.~\ref{tab:Re}).

\begin{table}
\begin{tabular}{p{1.8cm} | p{1.3cm} p{1.3cm} p{1.3cm} p{1.3cm} p{1.3cm} p{1cm}}
\hline \hline
$\Omega$ (rpm) & \hspace{0.1cm} 0  &  2 & 4 & 8 & 12 & 16  \\
\hline $Ro_f$ &  \hspace{0.1cm}  $\infty$ & 0.22 & 0.11 & 0.055 & 0.037 & 0.028 \\
\hline $k/K$ & \hspace{0.1cm} 0.48 & 0.79 & 0.89 & 0.95 & 0.94 & 0.97  \\
\hline $\gamma$ & \hspace{0.1cm} 0.17 & 0.15 & 0.06 & 0.05 & 0.04 & 0.10\\
\hline $L_ \perp$ (mm) & \hspace{0.1cm} 24 & 43 & 45 & 44 & 42 & 38 \\
\hline $Re$ & \hspace{0.1cm} 140 & 230 & 350 & 420 & 400 & 330\\
\hline $Ro $ & \hspace{0.1cm} $\infty$ & 0.30 & 0.20 & 0.13 & 0.087 & 0.068\\
\hline \hline
\end{tabular}
\caption{Flow parameters for the different rotation rates
$\Omega$: Rossby number based on the flap velocity $Ro_f$, rate of
turbulence $k/K$, inhomogeneity factor $\gamma$, horizontal
integral scale $L_{\perp}$, turbulent Reynolds number $Re$ and
Rossby number $Ro$. These figures are computed from the
Stereoscopic PIV data in the horizontal plane
(Fig.~\ref{fig:arene}(a)). See text for
definitions.}\label{tab:Re}
\end{table}

The three components of the velocity field ${\bf u}({\bf
x},t)=(u_x,u_y,u_z)$ are measured in a horizontal and a vertical
plane in the rotating frame (Fig.~\ref{fig:arene}(a)) using a
stereoscopic particle image velocimetry (PIV)
system.\cite{DaVis,Pivmat} The two regions of interest are
centered with respect to the arena of generators. It is a square
of about $14 \times 14$~cm$^2$ in a vertical plane along the
diagonal of the base of the tank, and a square of $12 \times
12$~cm$^2$ in a horizontal plane at mid-depth. The flow is seeded
with $10$~$\mu$m tracer particles and illuminated by a laser sheet
generated by a double $140$~mJ Nd:YAG pulsed laser mounted on the
rotating platform (Fig.~\ref{fig:arene}(a)). The illuminated flow
section is imaged with two double-buffer cameras aiming at the
laser sheet under different incidence angles. Images are taken
from above through the glass lid for the measurements in the
horizontal plane, and from two adjacent vertical sides of the tank
for the measurements in the vertical plane.

Each acquisition consists in 3\,600 quadruplets of images (one
pair per camera) recorded at 0.35~Hz with a $50$~ms time lag
between the two images of a given pair. The 3 velocity components
are computed in the two-dimensional measurement plane using
stereoscopic reconstruction. The cross-correlations are based on
interrogation windows of $32 \times 32$ pixels with $50\%$
overlap. The resulting 2D3C velocity fields are sampled on a grid
of $105 \times 105$ (resp. $80\times 80$)~vectors  with a spatial
resolution of $1.15$~mm (resp. $1.75$~mm) in the horizontal (resp.
vertical) plane.

\section{Local homogeneity and axisymmetry}
\label{sec:hom}

In this experiment, kinetic energy is injected by the generators
located around the region of interest, so an inward transport of
energy takes place from the generators to the center  of the
arena. An important feature of turbulence in this configuration is
the presence or not of a mean flow induced by the generators: this
indicates whether the transport of energy can be mainly attributed
to a reproducible flow or to turbulent fluctuations. This can be
addressed by performing a Reynolds decomposition of the velocity
field,
\begin{eqnarray}\label{eq:ReDecomp}
{\bf u}({\bf x},t)={\bf \bar{u}}({\bf x})+{\bf u}'({\bf x},t),
\end{eqnarray}
where ${\bf \bar{u}}({\bf x})$ is the time-averaged velocity field
and ${\bf u}'({\bf x},t)$ its turbulent part. From this
decomposition, we can compute the turbulent and total kinetic
energies, $k=\langle\overline{{\bf u}'({\bf x},t)^2}\rangle_{\bf
x}/2$ and $K=\langle\overline{{\bf u}({\bf x},t)^2}\rangle_{\bf
x}/2$ respectively, where $\langle \cdot \rangle_{\bf x}$ is a
spatial average over the horizontal region of interest. The
turbulence rate  $k/K$ is about $50\%$ in the non-rotating
experiment but rapidly increases up to 97\% as the rotation rate
$\Omega$ is increased (see Tab.~\ref{tab:Re}), indicating that,
under rotation, the turbulent structures are essentially
self-advected from the generators towards the center of the arena.
In the following we therefore focus on the turbulent component
${\bf u}'({\bf x},t)$ which dominates the flow in the rotating
case.

Although turbulence is necessarily inhomogeneous in this
configuration, with more energy near the generators than at the
center of the flow, we may expect a reasonable local homogeneity
in the measurement area, because of its small size (square of
about 13~cm side) compared to the distance to the generators
(33~cm from the center of the arena to the tip of the flaps).
Before investigating the scale-by-scale energy distribution and
energy transfers from spatially averaged two-point statistics, it
is therefore important to quantify the degree of homogeneity of
the flow. Since most of the energy is turbulent, we can quantify
the level of homogeneity by the spatial standard deviation of the
turbulent kinetic energy
\begin{eqnarray}\label{eq:inhomfact}
\gamma = \frac{\langle [k({\bf x}) - k] ^2 \rangle_{\bf
x}^{1/2}}{k},
\end{eqnarray}
with $k({\bf x}) = \overline{{\bf u}'({\bf x},t)^2} /2$ the
{\it local} time-averaged turbulent energy (such that
$k = \langle k({\bf x}) \rangle_{\bf x}$). This ratio is
given in Tab.~\ref{tab:Re}. It is smaller than $10\%$ for $\Omega>4$ rpm,
indicating a reasonable degree of homogeneity
in the region of interest.

A last single-point quantity of interest to characterize the
turbulence field in this configuration is the spatially averaged
velocity correlation tensor, $\langle \overline{u'_i
u'_j}\rangle_{\bf x}$ (the trace of this tensor is twice the
turbulence kinetic energy). For axisymmetric turbulence with
respect to $z$, one has $\langle \overline{u_x^{'2}} \rangle_{\bf
x} = \langle \overline{u_y^{'2}} \rangle_{\bf x} \neq \langle
\overline{u_z^{'2}} \rangle_{\bf x}$ (i.e. turbulence is isotropic
in the horizontal plane), with zero non-diagonal components. For
3D isotropic turbulence the three diagonal components are equal
(i.e., $\langle \overline{u'_i u'_j}\rangle_{\bf x} = \frac{2}{3}
k \, \delta_{ij}$). This tensor therefore characterizes the {\it
componentality} of turbulence, i.e. the isotropy with respect to
the velocity components. It must not be confused with the {\it
dimensionality} of turbulence, which characterizes the dependence
of the two-point velocity statistics with respect to orientation
of the separation vector joining the two points (investigated in
next section).

\begin{figure}
\centerline{\includegraphics[height=7cm]{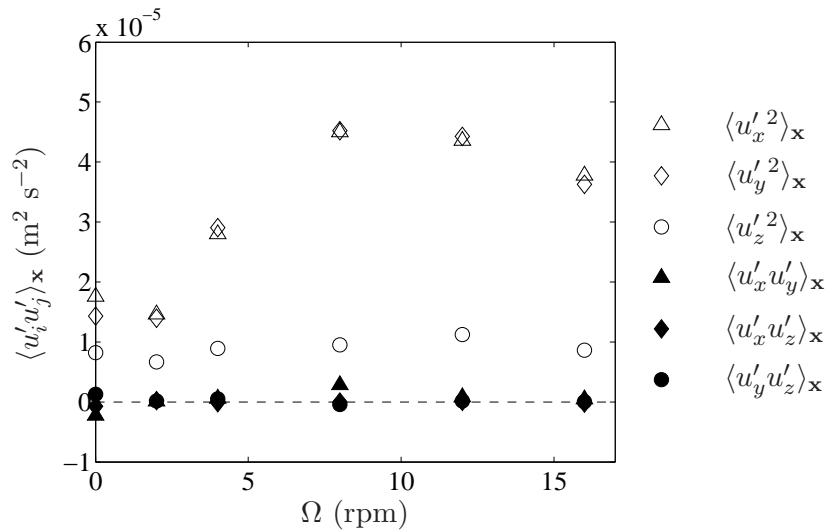}}
\caption{Components of the velocity correlation tensor $\langle
\overline{u_i' u_j'}\rangle_{\bf x}$ (with $(i,j)\in (x,y,z)$) as
a function of $\Omega$ averaged over the horizontal region of
interest. The non-diagonal components are nearly zero and $\langle
\overline{u_x'^2}\rangle_{\bf x} \simeq \langle
\overline{u_y'^2}\rangle_{\bf x}$, indicating statistically
axisymmetric turbulence. $\Omega=0, 2, 4, 8, 12, 16$~rpm
corresponds to turbulent Rossby number $Ro=\infty , 0.30, 0.20,
0.13, 0.087, 0.068$ respectively.}\label{fig:Rij}
\end{figure}

In Fig.~\ref{fig:Rij} we see that turbulence is nearly
axisymmetric, with $\langle \overline{u_x^{'2}} \rangle_{\bf x}
\simeq \langle \overline{u_y^{'2}} \rangle_{\bf x}$ to within 3\%
in the rotating case and 10\% in the non-rotating case, and with
the three non-diagonal components less than 10\% of the diagonal
components. As expected, turbulence is never isotropic, even in
the case $\Omega=0$, for which $\langle
\overline{u_x^{'2}}\rangle_{\bf x}\simeq\langle
\overline{u_y^{'2}}\rangle_{\bf x} \simeq 2\,\langle
\overline{u_z^{'2}}\rangle_{\bf x}$. This anisotropy originates
from the vertically-invariant forcing by the flaps, which induces
significantly weaker vertical velocity fluctuations than
horizontal ones. As the rotation rate $\Omega$ increases, $\langle
\overline{u_z^{'2}}\rangle_{\bf x}$ remains roughly constant,
whereas $\langle \overline{u_x^{'2}}\rangle_{\bf x}$ and $\langle
\overline{u_y^{'2}}\rangle_{\bf x}$ first increase with $\Omega$
and saturates beyond 8~rpm ($Ro \simeq 0.13$). At large $\Omega$,
the vertical kinetic energy represents about 10\% of the total
energy.

\section{Scale-by-scale energy distribution and transfers}

We now focus on the scale-by-scale energy distribution and energy
transfers. For this we must use two-point quantities: let us
consider two points A and B in the turbulent flow at positions
${\bf x}_A$ and ${\bf x}_B$. We define the mid-point position
${\bf X}=({\bf x}_A+{\bf x}_B)/2$ and the separation vector ${\bf
r}={\bf x}_B-{\bf x}_A$. Using cylindrical coordinates the
separation ${\bf r}$ writes ($r_\perp$,$\varphi$,$r_\parallel$),
with $r_\perp = (r_x^2 + r_y^2)^{1/2}$ and $r_\parallel = r_z$. In
homogeneous turbulence, all statistical averages are functions of
the separation vector ${\bf r}$ only. However, inhomogeneity plays
a key role in boundary forced experiments, and we thus consider
the inhomogeneous framework in which ensemble averages remain
functions of both ${\bf r}$ and ${\bf X}$.

The centered velocity increment for separation ${\bf r}$,
mid-point ${\bf X}$ and time $t$ is
\begin{eqnarray*}
\delta {\bf u'}({\bf X},{\bf r},t) &=& {\bf u}'_B({\bf X},{\bf r},t)-{\bf u}'_A({\bf X},{\bf r},t) \\
&=& {\bf u'}({\bf x}_B={\bf X} +
{\bf r}/2,t)-{\bf u'}({\bf x}_A = {\bf X} - {\bf r}/2,t).
\end{eqnarray*}
To perform an energy budget in this inhomogeneous context, we
first need to define a control region for the mid-point ${\bf X}$:
we consider only points A and B for which ${\bf X}$ lies in square
$S_X$ of side 40~mm centered in the PIV field. The relatively
small size of the square allows for the separation $|{\bf r}|$ to
be as large as $80$~mm with the two points A and B still lying in
the PIV field. The statistical averages are defined as an average
over time and over all the positions of ${\bf X}$ inside $S_X$. In
the following, the spatial average $\langle\,.\,\rangle_{{\bf X}}$
over ${\bf X} \in S_X$ will simply be denoted by $\langle\,.\,
\rangle$.

Defining the statistics from centered or non-centered
increments would be equivalent for homogeneous turbulence, and
most of the remainder of this section can be understood in this
framework. However, the use of centered increments plays a key
role in Sec.~\ref{sec:ikhm} where we discuss the scale-by-scale
energy budget: there is a balance at every scale ${\bf r}$ between
viscous dissipation, nonlinear transfers between different scales
(flux in ${\bf r}$ space), and advection of kinetic energy at
scale ${\bf r}$ through the boundaries of the domain $S_X$ (flux
in ${\bf X}$ space). The latter term is the source term of the
energy budget, which vanishes if turbulence is assumed to be
homogeneous.

\subsection{Energy distribution}
\label{sec:E}

\begin{figure}
\centerline{\includegraphics[width=11.5cm]{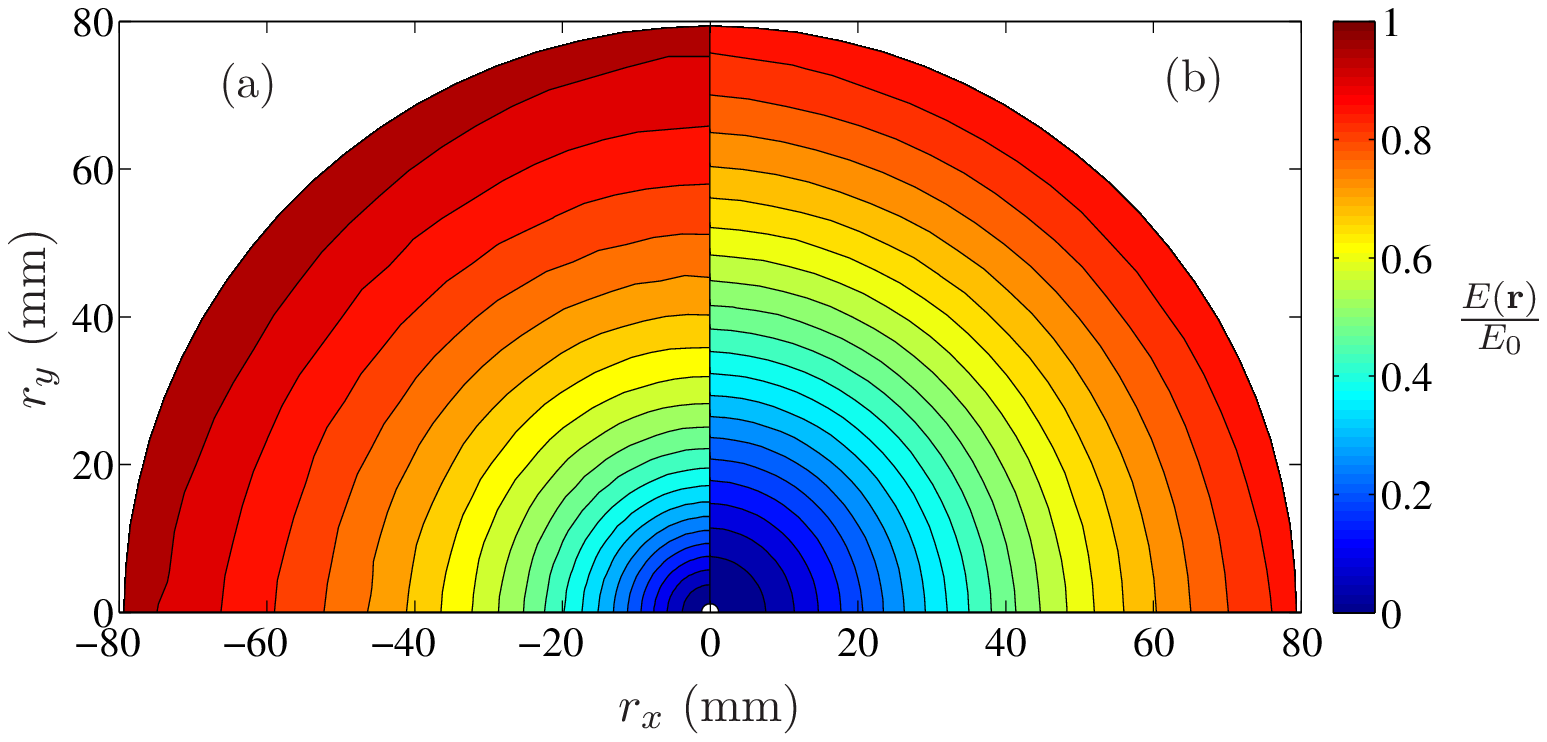}}
\vspace{0.2cm}
\centerline{\includegraphics[width=11.5cm]{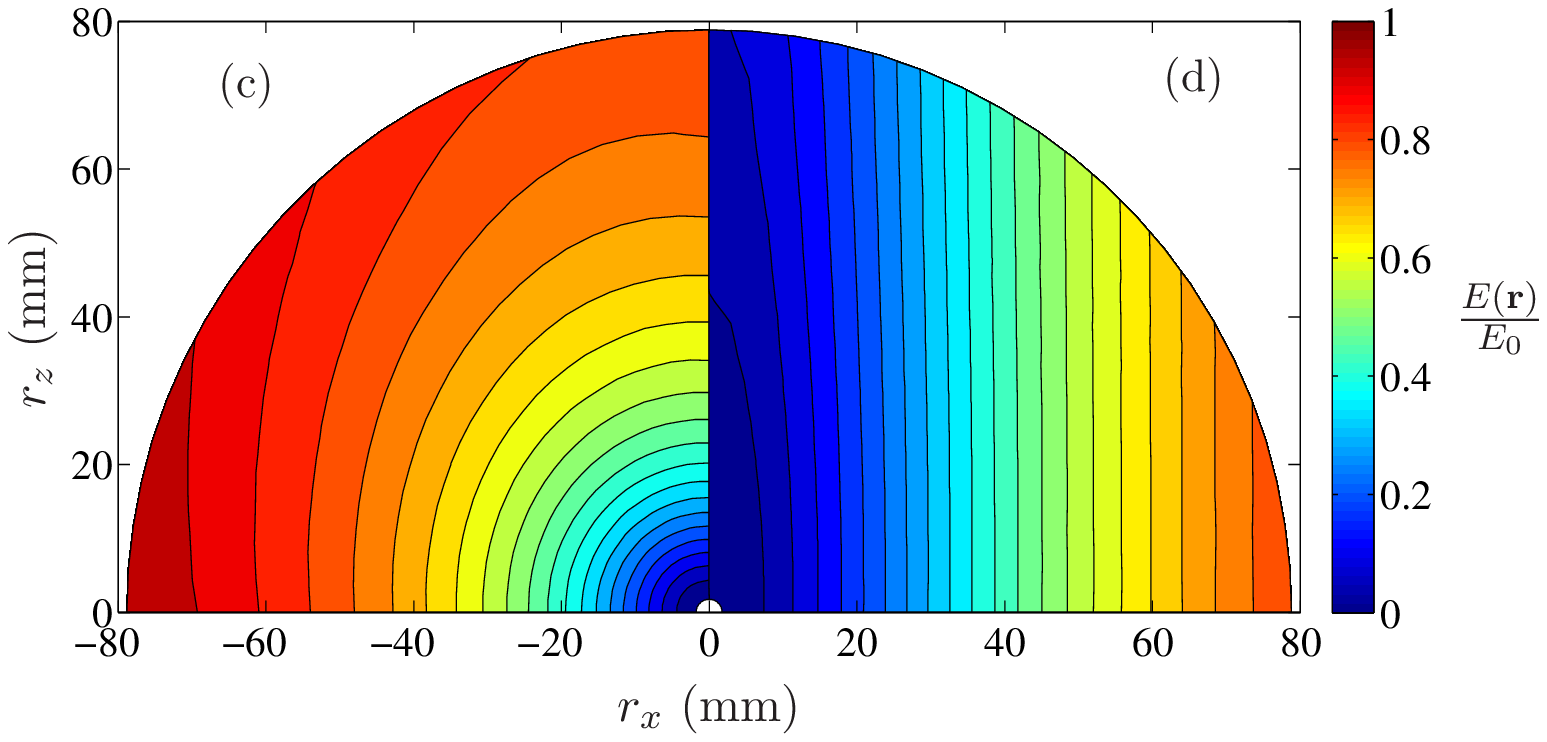}}
\caption{(Color online) Maps of the normalized energy distribution
$E({\bf r})/E_0$, (a,b) in the horizontal ($r_x$,$r_y$) plane and
(c,d) in the vertical ($r_x$,$r_z$) plane. (a) and (c) correspond
to the experiment with $\Omega=0$ ($Ro=\infty$), and (b) and (d)
to $\Omega=16$~rpm ($Ro=0.068$).}\label{fig:E}
\end{figure}

We characterize the distribution of the turbulent energy among
spatial scales by the anisotropic second-order structure function,
defined as the variance of the centered velocity
increments\cite{fn1}
\begin{equation}
E({\bf r})=\langle \overline{(\delta{\bf u'})^2}\rangle.
\label{eq:defE}
\end{equation}
The angular average of this quantity,
$\mathcal{E}(r)=1/(4\pi)\,\int_{\theta=0}^{\pi}\int_{\varphi=0}^{2\pi}E(r,\theta,\varphi)
\sin \theta d\theta d\varphi$, where $(r,\theta,\varphi)$ is the
usual spherical coordinate system, can be interpreted as the
energy contained in eddies of size $r$ or less, provided that $r$
is larger than the dissipative
scale.\cite{DavidsonBook2004,DavidsonBook2013} For isotropic
turbulence, $E({\bf r})=\mathcal{E}(r)$ therefore directly
measures the cumulative energy from $0$ to $r$. For anisotropic
turbulence, $E({\bf r})$ contains in addition key information on
the anisotropic distribution of energy among eddies of
characteristic horizontal and vertical scale given by $r_\perp$
and $r_\parallel$ respectively. For isotropic turbulence, the
isosurfaces of $E({\bf r})$ are spherical, while for axisymmetric
turbulence about the vertical they are invariants with respect to
rotations around the $r_z$~axis. Two-dimensional turbulence would
give exactly cylindrical iso-$E({\bf r})$ (invariant by
translation along $r_z$), which is a special case of axisymmetric
turbulence.

Figure~\ref{fig:E} shows the maps of the normalized energy
distributions $E({\bf r})/E_0$ in the horizontal $(r_x,r_y)$ and
vertical $(r_x,r_z)$ planes for $\Omega=0$ and 16~rpm, with
$E_0=\langle\overline{{\bf u}'^2_A+{\bf u}'^2_B}\rangle_{\bf X}$
taken at ${\bf r}=r_{\rm max}{\bf e}_\perp$ and $r_{\rm
max}=80$~mm the maximum separation.\cite{fn2} In
Figs.~\ref{fig:E}(a,b), the iso-contours of $E({\bf r})$ are
nearly circular in the horizontal plane, both without and with
rotation, indicating the good level of two-point axisymmetry of
turbulence. For the largest horizontal scales considered here
($|{\bf r}|=80$~mm), $E/E_0$ reaches $0.98$ for $\Omega=0$,
indicating that nearly all turbulent energy is contained in the
range of scales of interest, whereas it reaches $0.89$ only at
$\Omega=16$~rpm, indicating that structures larger than the
maximum available scale still contain energy. This is a first
indication of the emergence of large horizontal structures in the
presence of rotation. This effect can be further quantified by the
horizontal integral scale,
\begin{eqnarray}
L_{\perp}=\int_0^{r^*} \mathcal{C}(r_\perp) dr_\perp,
\label{eq:Lperp}
\end{eqnarray}
with $\mathcal{C}(r_\perp)=1/(2\pi)\int_{0}^{2\pi} C({\bf
r}_\perp) d\varphi$, and $C({\bf r}_\perp) = 2\,\langle {\bf u'}_A
\cdot {\bf u'}_B \rangle_{\bf X} / E_0$ the two-point correlation
function. The conventional definition is such that $r^* = \infty$,
but using here a finite truncation at $r^*$, chosen such that
$\mathcal{C}(r^*) = 0.25$, is necessary because
$\mathcal{C}(r_\perp)$ does not reach $0$ at the maximum available
scale $r_\perp=80$~mm. Values of $L_{\perp}$ are given in
Tab.~\ref{tab:Re}. In the absence of rotation, $L_\perp \simeq
24$~mm, which corresponds to the characteristic size of the
turbulent fluctuations generated by the flaps. As $\Omega$ is
increased, $L_\perp$ grows by nearly a factor 2 compared to the
non-rotating case, confirming the generation of large-scale
structures.

\begin{figure}
\centerline{\includegraphics[height=6.5cm]{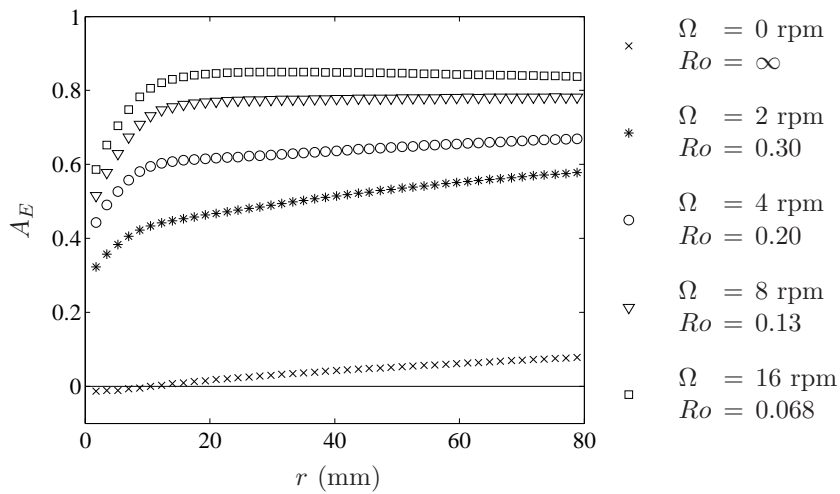}}
\caption{Scale dependent anisotropy factor $A_E(r)$
[Eq.~(\ref{eq:ae})] of the energy distribution as a function of
$r$ for different rotation rates ($A_E = 0$ for 3D isotropic
turbulence and $A_E = 1$ for 2D3C turbulence).} \label{fig:CE}
\end{figure}

We now turn to the energy distribution in the vertical plane
(Figs.~\ref{fig:E}c,d). $E({\bf r})$ is anisotropic both with and
without rotation, with a trend towards vertical elongation of the
contour lines at large scales. This anisotropy is however weak at
$\Omega=0$ and affects preferentially the large scales: this is a
direct consequence of the vertical invariance of the forcing
device, which creates a nearly 2D flow at large scale carrying
small-scale 3D fluctuations. The anisotropy  is much more
pronounced in the presence of rotation and persists down to the
smallest scales, indicating a trend towards quasi-2D turbulence.
This scale dependence of the anisotropy can be quantified by the
ratio
\begin{equation}
A_E(r)=\frac{E_\perp(r)-E_\parallel(r)}{E_\perp(r)+E_\parallel(r)},
\label{eq:ae}
\end{equation}
with $E_\perp(r)=E(r_x=r,r_z=0)$ and
$E_\parallel(r)=E(r_x=0,r_z=r)$. This ratio is zero for 3D
isotropic turbulence and 1 for 2D turbulence. The plot of $A_E(r)$
in Fig.~\ref{fig:CE} shows a growth of anisotropy with $r$ at all
rotation rates. This growth is weak for $\Omega=0$ (with $A_E$
increasing from -0.01 to 0.1), indicating that the 2D nature of
the forcing has a weak influence at these scales in the center of
the tank. The anisotropy is much stronger when $\Omega \neq 0$:
$A_E(r)$ first grows rapidly from $r=0$ to $r \sim 10$~mm before
saturating. For the largest available rotation rate,
$\Omega=16$~rpm, turbulence is nearly 2D for $r>10$~mm, with
$A_E(r) \simeq 0.85$, but remains significantly 3D at smaller
scales.

\subsection{Energy transfers}
\label{sec:E}

We now consider the scale-by-scale energy transfers, defined
from third-order moments of velocity increments. We start from
the K\'{a}rm\'{a}n-Horwarth-Monin equation for time-dependent
homogeneous (but not
necessarily isotropic) turbulence\cite{Monin1975,FrischBook}
\begin{equation}
\frac{1}{4}\partial_t E({\bf r},t) = -\Pi({\bf r},t) + \frac{1}{2}
\nu\nabla_{\bf r}^2 E - \epsilon,\label{eq:KHME}
\end{equation}
where $E({\bf r},t) = \langle (\delta {\bf u}')^2 \rangle_{{\bf
X},E}$, and
\begin{equation}
\Pi({\bf r},t)=\frac{1}{4}{\boldsymbol \nabla}_{\bf r}\cdot
\langle{(\delta{\bf u'})^2\delta{\bf u'}}\rangle_{{\bf X},E}
\label{eq:defPi}
\end{equation}
is the energy transfer term in scale-space (with ${\boldsymbol
\nabla}_{\bf r}\cdot$ the divergence with respect to the vector
separation ${\bf r}$), and $\epsilon = \nu \langle {(\partial_i
u_j' +\partial_j u_i')^2} \rangle/2$ the instantaneous energy
dissipation rate. Here the brackets $\langle \cdot \rangle_{{\bf
X},E}$ represent spatial and ensemble average. Similarly to the
angular average $\mathcal{E}(r)$ of $E({\bf r})$, which represents
the cumulative energy from scale $0$ to $r$, the angular average
$\mathcal{P}(r)=(4\pi)^{-1}\,\int_{\theta=0}^{\pi}\int_{\varphi=0}^{2\pi}\Pi(r,\theta,\varphi)
\sin \theta d\theta d\varphi$ can be interpreted as the energy
flux from scales smaller than $r=|{\bf r}|$ towards scales larger
than $r$. For isotropic turbulence, the sign of $\Pi({\bf
r})=\mathcal{P}(r)$ therefore gives the direction of the energy
cascade, forward if $\Pi({\bf r})<0$ and inverse if $\Pi({\bf
r})>0$. In the inhomogeneous case, additional terms corresponding
to advection of energy between different regions of the turbulent
flow appear in Eq.~(\ref{eq:KHME}). In the absence of body forces,
which are not relevant in our experiment, advection of energy from
outside the control domain is the only source term to sustain
stationary turbulence: we will come back to this point in
Sec.~\ref{sec:ikhm}.

In the following we focus on stationary turbulence, and we take
$\partial_t E = 0$ in Eq.~(\ref{eq:KHME}). The ensemble average
$\langle \cdot \rangle_E$ can be therefore replaced by a temporal
average, which we denote as $\overline·$. For axisymmetric
turbulence, it is convenient to decompose the flux
(\ref{eq:defPi}) into its perpendicular (horizontal) and parallel
(vertical) contributions,
\begin{eqnarray}
\Pi({\bf r}) &=& \Pi_\perp ({\bf r}) + \Pi_\parallel ({\bf r}),\label{eq:piperppipara} \nonumber \\
&=& \frac{1}{4} {\boldsymbol \nabla}_\perp \cdot
\langle\overline{(\delta{\bf u'})^2\delta{\bf u'}_\perp}\rangle +
\frac{1}{4} \nabla_\parallel \langle\overline{(\delta{\bf
u'})^2\delta u'_\parallel}\rangle,
\end{eqnarray}
with ${\boldsymbol \nabla}_\perp = {\bf e}_x \partial_{r_x} + {\bf
e}_y \partial_{r_y}$ and $\nabla_\parallel = \partial_{r_z}$. We
focus in the following on pure horizontal separations by
setting $r_\parallel \equiv r_z= 0$, and we perform an azimuthal
average over $\varphi$ to improve the statistics. Both contributions
from Eq.~(\ref{eq:piperppipara}) are then functions of the
horizontal separation $r_\perp$ only.

For strictly 2D turbulence, vertical invariance implies
$\Pi_\parallel=0$. The vertical flux $\Pi_\parallel$ cannot be
measured here, because we cannot access the vertical derivative
$\nabla_\parallel$ from measurements in the horizontal plane.  In
principle one could use the data in the vertical plane, but we
found significant departure from axisymmetry for third-order
quantities (although second-order quantities are found nearly
axisymmetric, as shown in Fig.~\ref{fig:E}a). This lack of
axisymmetry can be circumvented by performing an azimuthal average
with respect to $\varphi$, which is possible in the horizontal
plane only.

\begin{figure}
\centerline{\includegraphics[width=12cm]{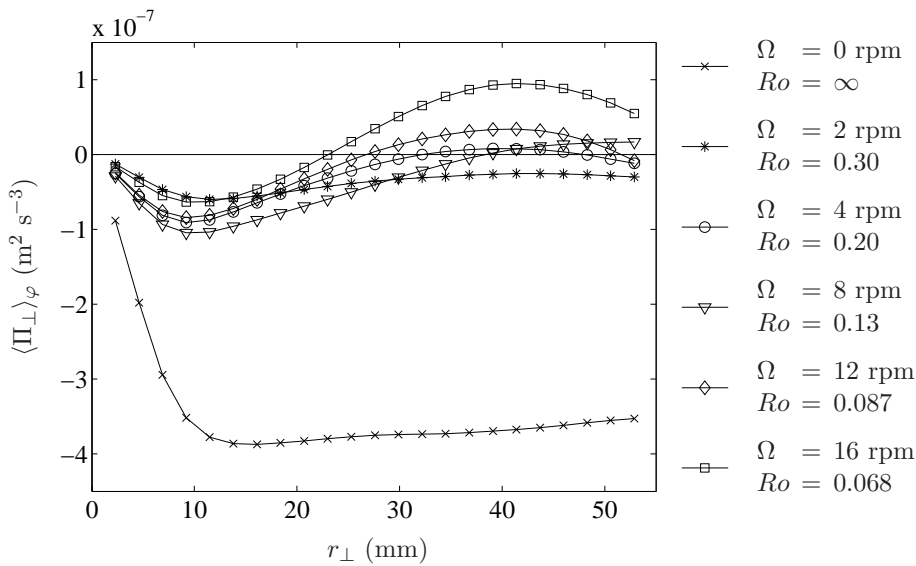}}
\caption{Azimuthal average of the horizontal energy flux,
$\langle\Pi_{\perp}\rangle_\varphi$, as a function of horizontal
separation $r_\perp$ and for various rotation rates $\Omega$.
These data are computed from measurements in the horizontal plane.
A negative value (resp. positive) corresponds to a direct (resp.
inverse) energy transfer.}\label{fig:divdu2du}
\end{figure}

For the non-rotating experiment, Fig.~\ref{fig:divdu2du} shows
that the horizontal flux $\Pi_\perp(r_\perp)$ is negative at all
scales, as expected for a direct energy cascade from large to
small scales, with $\Pi_\perp \rightarrow 0$ as $r_\perp
\rightarrow 0$ in the viscous range. \cor{The observed $10\%$
decrease in $|\Pi_\perp(r_\perp)|$ for $r_\perp$ beyond 15~mm
might be due to the vertical invariance of the forcing device: the
large scales are slightly 2D (see Fig.~\ref{fig:CE}), which may
enhance the inverse energy transfers and reduce the direct ones.
However, as discussed in section~\ref{sec:ikhm}, boundary-driven
flows display some inhomogeneity at large scale, which challenges
the interpretation of $\Pi_\perp$ in terms of pure scale-by-scale
energy transfers. Therefore, we do not elaborate more on the
behavior of $\Pi_\perp$ at scales larger than $r_\perp>40$~mm (see
details in section~\ref{sec:ikhm}).}

For increasing rotation rates, $\Pi_\perp(r_\perp)$ strongly
decreases at intermediate scales in absolute value, and eventually
a change of sign is observed at large scales beyond 4~rpm ($Ro
\simeq 0.2$). This indicates the onset of an inverse energy
cascade, which spreads towards smaller scales as $\Omega$
increases. Remarkably, the double cascade persists even at the
largest rotation rate ($\Omega=16$~rpm, $Ro \simeq 0.068$), with
the coexistence of an inverse flux ($\Pi_\perp > 0$) at large
scales and a direct flux ($\Pi_\perp < 0$) at small scales. This
implies that, on average, energy must be supplied at intermediate
scales (of the order of 25~mm for $\Omega=16$~rpm): we return to
this point in Sec.~\ref{sec:ikhm}.

\subsection{Horizontal transfers of horizontal and vertical energy}

\begin{figure}
\centerline{\includegraphics[height=14cm]{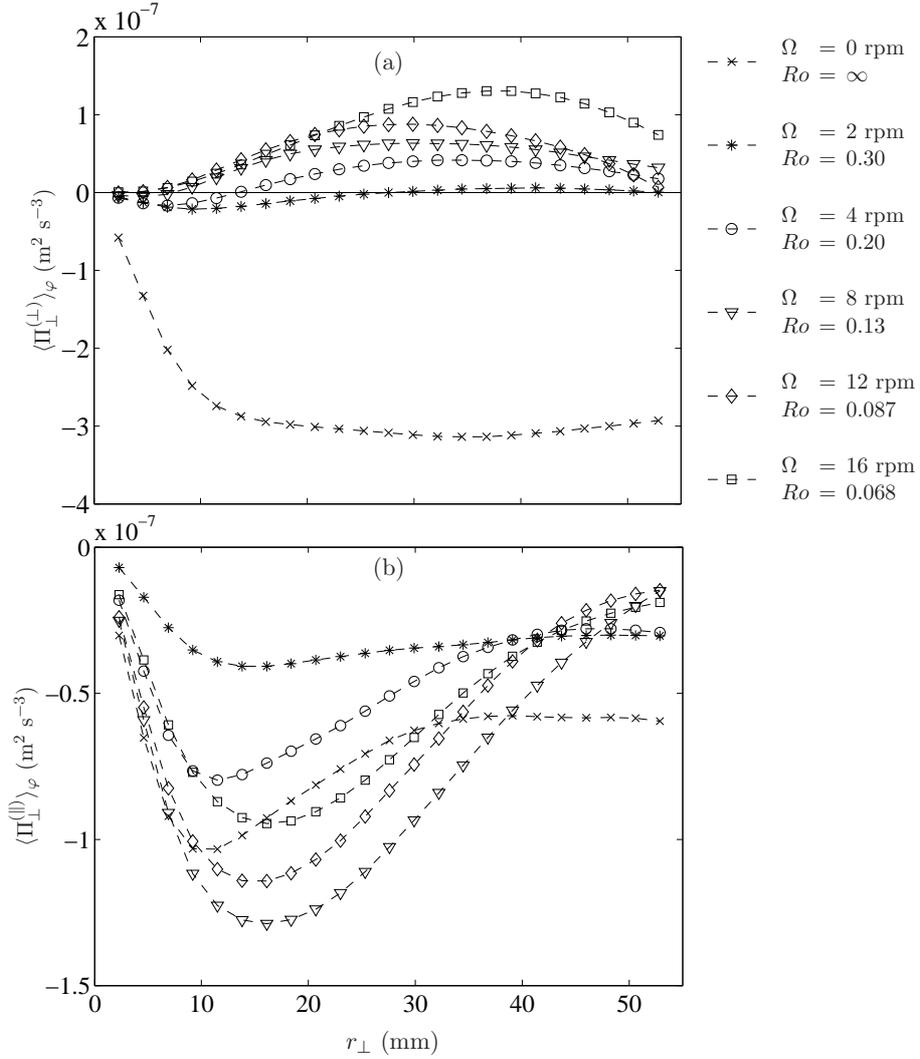}}
\caption{Horizontal flux of (a) horizontal energy,
$\Pi_{\perp}^{(\perp)}(r_\perp)$, and (b) vertical energy,
$\Pi_{\perp}^{(\parallel)}(r_\perp)$, at various rotation rates
$\Omega$. A negative flux corresponds to a direct energy transfer
(from large to small scales), whereas a positive flux corresponds
to an inverse energy transfer.}\label{fig:divduz2du}
\end{figure}

The energy flux $\Pi_\perp(r_\perp)$ contains both the horizontal
flux of horizontal energy, $(\delta {\bf u'}_\perp)^2$, and the
horizontal flux of vertical energy, $(\delta u'_\parallel)^2$. To
get further insight into the double cascade observed in
Fig.~\ref{fig:divdu2du}, we decompose $\Pi_\perp$ as follows
\begin{eqnarray}
\Pi_{\perp}(r_\perp) &=& \Pi_\perp^{(\perp)}(r_\perp) + \Pi_\perp^{(\parallel)}(r_\perp) \label{eq:decompPi}\\
&=& \frac{1}{4} {\boldsymbol \nabla}_{\perp} \cdot
\langle\overline{(\delta{\bf u'}_\perp)^2\delta{\bf
u'}_\perp}\rangle + \frac{1}{4} {\boldsymbol \nabla}_{\perp} \cdot
\langle\overline{(\delta u'_\parallel)^2\delta{\bf
u'}_\perp}\rangle.
\label{eq:Pipp}
\end{eqnarray}
These two contributions are shown in Fig.~\ref{fig:divduz2du}.
Interestingly, we observe that $\Pi_\perp^{(\parallel)}$ remains
negative at all rotation rates, indicating that  vertical energy
is always transferred from large to small horizontal scales,
whereas $\Pi_\perp^{(\perp)}$ becomes positive as the rotation
rate is increased, a signature of the onset of an inverse cascade
for the horizontal energy. In the non-rotating case, this negative
flux $\Pi_\perp^{(\parallel)}$ is compatible with the classical
direct cascade framework of 3D turbulence. By contrast, in the
presence of rotation, for the scales at which the inverse cascade
of $(\delta {\bf u'}_\perp)^2$ is observed, the direct cascade of
$(\delta u'_\parallel)^2$ is reminiscent of the behavior of a
passive scalar advected by a two-dimensional flow: the stretching
and folding of the vertical velocity by the horizontal flow
produce small scales through filamentation, inducing a direct
horizontal cascade of vertical
velocity.\cite{Yaglom1949,Monin1975,Tabeling2002} We provide in
the next section further assessment of this picture.

Figure~\ref{fig:divduz2du} also indicates that the horizontal flux
of vertical energy $\Pi_\perp^{(\parallel)}$ is a significant
contribution to $\Pi_\perp$ for all rotation rates. For the low
rotation rate $\Omega = 2$~rpm ($Ro = 0.3$), although a
significant inverse cascade already takes place at large scale for
the horizontal energy ($\Pi_\perp^{(\perp)} > 0$), it is hidden by
a stronger direct cascade of vertical energy
($\Pi_\perp^{(\parallel)} < 0$). This results in an overall direct
cascade of total energy ($\Pi_\perp < 0$). For larger rotation
rates, the inverse cascade of horizontal energy becomes dominant,
eventually leading to $\Pi_\perp^{(\perp)} > 0$ at all scales for
$\Omega \geq 12$~rpm. The crossover scale separating the direct
and inverse cascades of horizontal energy rapidly decreases as
$\Omega$ increases, going from $\sim 30$~mm for $\Omega=2$~rpm to
zero for $\Omega \geq 12$~rpm (and then $\Pi_\perp^{(\perp)} > 0$
over the whole range of scales).

\section{Scale-by-scale energy budget}
\label{sec:ikhm}

\subsection{Inhomogeneous K\'{a}rm\'{a}n-Howarth-Monin equation}
\label{sec:ikhme}

To provide a physical interpretation for the sign of the
scale-by-scale energy flux $\Pi({\bf r})$, we must
describe carefully the energy input in the experiment, and in
particular its scale dependence. In numerical simulations of
homogeneous stationary turbulence, this source term usually
originates from a random body force acting on a narrow range of
scales. By contrast, here the fluid motion is driven by moving
solid boundaries, so the energy injection in a given control
volume away from the forcing must originate from the transport of
energy through the surface delimiting the control volume. Since a
non-trivial stationary state cannot be described by the
homogeneous non-forced KHM equation~(\ref{eq:KHME}), which
contains no source term, we must consider explicitly the effects
of the inhomogeneities in the following.

We consider here the inhomogeneous generalization of the KHM
equation proposed by Hill.\cite{Hill2002} We briefly recall
the derivation of this equation, with the addition of the Coriolis
force. Let us start from the incompressible Navier-Stokes (NS)
equation in the rotating frame
\begin{eqnarray}
\partial_t{\bf u}+({\bf u}\cdot{\nabla}){\bf u} = -{\bf \nabla}p-2{\boldsymbol \Omega}\times{\bf u}+\nu{\bf \nabla}^2{\bf u},
\label{eq:NS}
\end{eqnarray}
with $p$ the pressure modified by the centrifugal force and
normalized by the fluid density.
Taking the difference between points ${\bf x}_B = {\bf X}+{\bf r}/2$ and ${\bf x}_A = {\bf X}-{\bf r}/2$ and taking the scalar
product with $\delta{\bf u} = {\bf u}_B - {\bf u}_A$ yield
\begin{eqnarray}
\begin{split}
\partial_t(\delta{\bf u})^2+{\boldsymbol \nabla}_{{\bf r}}\cdot (\delta {\bf u})^2\delta{\bf u} = & 2\nu{\boldsymbol \nabla}_{\bf r}^2(\delta{\bf u})^2-4\tilde{\epsilon}\\
& + {\boldsymbol \nabla}_{{\bf X}}\cdot \left[ -(\delta {\bf u})^2\tilde{\bf u}-2\delta p\delta{\bf u}+\frac{\nu}{2}{\boldsymbol \nabla}_{{\bf X}}\left( (\delta {\bf u})^2-8\tilde{{\boldsymbol \tau}} \right) \right],
\end{split}
\label{eq:KHM_inhom_nomoy}
\end{eqnarray}
with $\delta p = p_B - p_A$. Quantities with $\tilde·$
denote the average between the two points: $\tilde{\bf u}=({\bf
u}_A + {\bf u}_B)/2$, $\tilde{p} = (p_A + p_B)/2$, and
$\tilde{\epsilon}=({\epsilon}_A+{\epsilon}_B)/2$, with
$\epsilon=\frac{\nu}{2}(\partial_{j}u_i+\partial_{i}u_j)^2$ the
local energy dissipation rate. The last term of the equation
involves the velocity correlation tensor $\tilde{{\boldsymbol
\tau}}=({\boldsymbol \tau}_A+{\boldsymbol \tau}_B)/2$, with
${\boldsymbol \tau}_{ij}=u_i u_j$. Importantly, all the terms in
Eq.~(\ref{eq:KHM_inhom_nomoy}) are functions of $({\bf X}, {\bf
r}, t)$, and the nonlinear term splits into a scale-to-scale
transfer term (divergence w.r.t. separation ${\bf r}$) and a
transport term (divergence w.r.t. mid-point ${\bf X}$).

We consider both the spatial average $\langle \cdot \rangle_{\bf
X}$ over a control volume $V_X$ and the ensemble average $\langle
\cdot \rangle_E$ of Eq.~(\ref{eq:KHM_inhom_nomoy}). Using the
divergence theorem to express the inhomogeneous terms as a flux
through the closed surface $S_X$ delimiting the control volume
$V_X$, we obtain
\begin{equation}
\partial_t \langle(\delta{\bf u})^2\rangle_{{\bf X},E} + {\boldsymbol \nabla}_{{\bf r}}\cdot \langle(\delta {\bf u})^2\delta{\bf u}\rangle_{{\bf X},E}=2\nu{\boldsymbol \nabla}_{\bf r}^2\langle(\delta{\bf u})^2\rangle_{{\bf X},E}-4\langle\tilde{\epsilon}\rangle_{{\bf X},E}+\Phi_{\rm inh}({\bf r}),
\label{eq:KHM_inhom_moyX}
\end{equation}
where the flux term writes
\begin{equation}
\Phi_{\rm inh}({\bf r})=\frac{1}{V_X} \oint_{S_X} \left(
-\langle(\delta {\bf u})^2\tilde{\bf u}\rangle_E - 2\langle\delta
p \, \delta{\bf u}\rangle_E + \frac{\nu}{2}{\boldsymbol
\nabla}_{{\bf X}}\langle (\delta {\bf u})^2-8\tilde{{\boldsymbol \tau}} \rangle_E
\right) \cdot d{\bf S_X}. \label{eq:phiinh}
\end{equation}
The unit vector $d{\bf S_X}$ is directed outward of the control
volume by convention. In the scale-by-scale
budget~(\ref{eq:KHM_inhom_moyX}), the energy input (or output) at
a given scale ${\bf r}$ is ensured by the term $\Phi_{\rm
inh}({\bf r})$, which originates from the inhomogeneities in the
pressure and velocity statistics. For homogeneous turbulence, one
has $\langle\tilde{\epsilon}\rangle_{{\bf
X},E}=\langle\epsilon\rangle_{{\bf x},E}$ and $\Phi_{\rm inh}=0$,
so Eq.~(\ref{eq:KHM_inhom_moyX}) becomes the usual KHM equation
(\ref{eq:KHME}).

The flux term (\ref{eq:phiinh}) contains
three contributions:
\begin{itemize}

\item[(i)] $\oint-\langle(\delta {\bf u})^2\tilde{\bf u}\rangle_E
\cdot d{\bf S_X}$ is the flux of cumulative energy $(\delta {\bf
u})^2$ through the surface $S_X$ due to advection by the velocity
$\tilde{\bf u}=({\bf u}_A+{\bf u}_B)/2$. It is positive when
$\tilde{\bf u}$ is directed into the control volume. Note that
this term takes a simple form in the classical configuration of a
wind-tunnel: the transport velocity $\tilde{\bf u}$ is essentially
replaced by the uniform mean velocity ${\bf U_0}$, and the energy
flux per unit surface becomes $\oint- \langle(\delta {\bf
u})^2\rangle_E \,{\bf U_0} \cdot d{\bf S_X}$. The inward flux of
$(\delta {\bf u})^2$ through the upstream face of the control
volume is larger than the outward flux through the downstream
face, hence a net flux of kinetic energy into the control volume,
which is dissipated at the same rate by viscosity. By contrast,
the time-averaged velocity is negligible in the present experiment
when $\Omega>0$, and energy input in the control volume proceeds
through advection of turbulent kinetic energy by the turbulent
velocity itself.

\item[(ii)] The term $\oint- 2\langle\delta p\, \delta{\bf
u}\rangle_E\cdot d{\bf S_X}$ originates from the work of the
pressure force through the boundary of the control volume. This
term cannot be measured experimentally. However, under the
assumption of local axisymmetry and homogeneity at small scale,
this term is expected to be much smaller than the advection term
for scales much smaller than the characteristic scale of
inhomogeneity (see Appendix~\ref{sec:app}).

\item[(iii)] The diffusion term in Eq.~(\ref{eq:phiinh}) is weak
because it corresponds to derivatives with respect to ${\bf X}$,
which are negligible compared to derivatives with respect to ${\bf
r}$ at small scales for weakly inhomogeneous turbulence (see
Appendix~\ref{sec:app}).

\end{itemize}

We stress the fact that the present inhomogeneous KHM equation has
a clear interpretation for scales smaller than the characteristic
scale of inhomogeneity, for which the contribution from the
pressure can be neglected. This is because the two-point
velocity-pressure correlation can be written either as a
divergence with respect to ${\bf X}$ or with respect to ${\bf r}$,
so the interpretation of $\Pi$ as the only scale-by-scale transfer
term becomes questionable when the velocity-pressure correlation
is significant, i.e. for large (inhomogeneous) scales. In the
following we show that interesting modifications of the energy
transfers by global rotation occur in the range of scales for
which the velocity-pressure correlation is indeed negligible.

\subsection{Scale-by-scale budget for 2D3C flows}

For rapid global rotation, the flow becomes weakly dependent on
the vertical coordinate far from the horizontal boundaries. In the
following we assume a purely 2D3C velocity field in the bulk of
the flow, and write separate equations for the evolution of the
horizontal and vertical energies. In this 2D
limit, the same analysis as in Sec.~\ref{sec:ikhme}, but performed
here on the horizontal components of Eq.~(\ref{eq:NS}) only,
yields
\begin{eqnarray}
\begin{split}
\partial_t \langle(\delta{\bf u}_{\perp})^2\rangle_{{\bf X},E} + {\boldsymbol \nabla}_{{\bf r}}\cdot \langle(\delta {\bf u}_{\perp})^2\delta{\bf u}_{\perp}\rangle_{{\bf X},E}=2\nu{\boldsymbol \nabla}_{\bf r}^2\langle(\delta{\bf u}_{\perp})^2\rangle_{{\bf X},E}-4\langle\tilde{\epsilon}_{\perp}^{(\perp)}\rangle_{{\bf X},E}+\Phi_{\rm inh}^{(\perp)}({\bf r}),
\label{eq:KHM_inhom_moyX_2D3C}
\end{split}
\end{eqnarray}
with
\begin{equation}
\Phi_{\rm inh}^{(\perp)}({\bf r})=\frac{1}{V_X} \oint_{S_X} \left(
-\langle(\delta {\bf u}_{\perp})^2\tilde{\bf u}_{\perp}\rangle_E - 2\langle\delta
p \, \delta{\bf u}_{\perp}\rangle_E + \frac{\nu}{2}{\boldsymbol
\nabla}_{{\bf X}}\langle (\delta {\bf u}_{\perp})^2-8\tilde{{\boldsymbol \tau}}_{\perp} \rangle_E
\right) \cdot d{\bf S_X} \, , \label{eq:phiinh_2D3C}
\end{equation}
where the subscript $\perp$ in $\delta {\bf u}_{\perp}$,
$\tilde{\bf u}_{\perp}$, and $\tilde{{\boldsymbol \tau}}_{\perp}$
indicates that only the horizontal velocity components are
considered, and the $\tilde·$ indicates that the quantity is a
mid-point average. This equation is an inhomogeneous version of
the KHM equation for the horizontal flow only. Similarly, from the
vertical component of the Navier-Stokes equation, one can also
compute the budget for the vertical energy,
\begin{eqnarray}
\begin{split}
\partial_t \langle(\delta{ u}_{\parallel})^2\rangle_{{\bf X},E} + {\boldsymbol \nabla}_{{\bf r}}\cdot \langle(\delta { u}_{\parallel})^2\delta{\bf u}_{\perp}\rangle_{{\bf X},E}=2\nu{\boldsymbol \nabla}_{\bf r}^2\langle(\delta{u}_{\parallel})^2\rangle_{{\bf X},E}-4\langle\tilde{\epsilon}_\perp^{(\parallel)}\rangle_{{\bf X},E}+\Phi_{\rm inh}^{(\parallel)}({\bf r}),
\label{eq:Yaglom_inhom_2D3C}
\end{split}
\end{eqnarray}
with
\begin{equation}
\Phi_{\rm inh}^{(\parallel)}({\bf r})=\frac{1}{V_X} \oint_{S_X} \left(
-\langle(\delta { u}_{\parallel})^2\tilde{\bf u}_{\perp}\rangle_E  + \frac{\nu}{2}{\boldsymbol
\nabla}_{{\bf X}} \langle (\delta {u}_{\parallel})^2-8\tilde{\tau}_{\parallel} \rangle_E
\right) \cdot d{\bf S_X} \, , \label{eq:phiinh_Yaglom}
\end{equation}
where the subscript $\parallel$ refers to the vertical component
of the velocity. This equation is an inhomogeneous generalization
of Yaglom's equation for a passive scalar
field:\cite{Yaglom1949,Monin1975} the tracer $u_{\parallel}$
enters and leaves the control volume through advection by the
horizontal velocity $\tilde{\bf u}_{\perp}$. Inside the control
volume, nonlinearities transfer the vertical energy between
different scales ${\bf r}$ through stretching and folding by the
horizontal field, and viscosity damps the strong gradients created
by these processes. We stress the fact that
Eq.~(\ref{eq:Yaglom_inhom_2D3C}) does not involve pressure: all
the terms in this equation can be accessed through stereo-PIV
measurements in a horizontal plane.

\subsection{Experimental assessment of the horizontal and vertical kinetic energy budget}

In the following we focus on the highest rotation rate,
$\Omega=16$~rpm ($Ro = 0.068$), for which we expect the turbulent
flow to reach a quasi-2D3C state, so that we can apply the $\perp$
vs. $\parallel$ decomposition of the inhomogeneous KHM equation
derived above. Note that in this case the ensemble-averaged flow
is negligible (see Tab.~\ref{tab:Re}), so we simply consider ${\bf
u}= {\bf u}'$.

We consider for the control domain a centered square in the square
PIV field, with a maximum separation $r_\perp=60$~mm. One can
think about the corresponding control volume as a parallelepiped
of arbitrary vertical length, with zero fluxes through the top and
bottom boundaries. Because turbulence is stationary in the
experiment, we replace the ensemble average $\langle \cdot
\rangle_E$ by a temporal one, which we denote $\overline·$.

\begin{figure}
\centerline{\includegraphics[height=13cm]{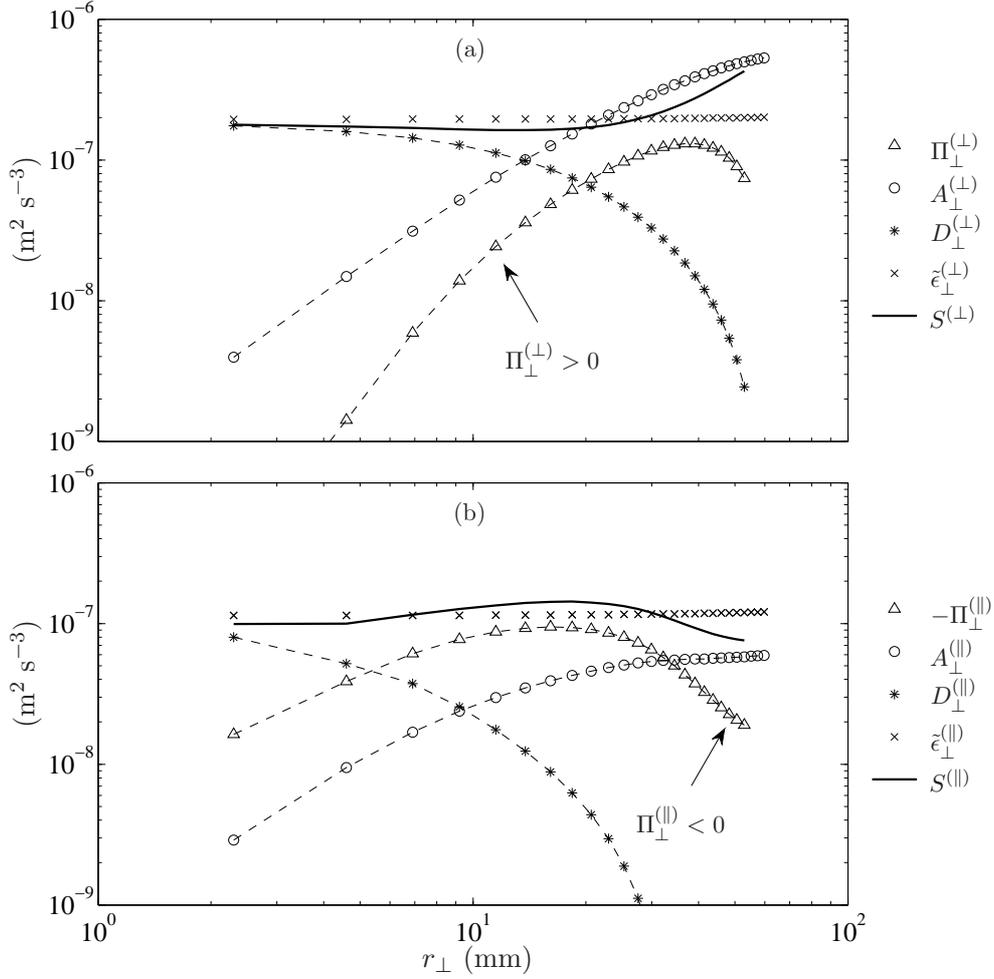}}
\caption{Scale-by-scale energy budget in the horizontal plane, for
(a) the horizontal kinetic energy $(\delta {\bf u}'_\perp)^2$, and
(b) the vertical kinetic energy $(\delta {u}'_\parallel)^2$\cor{:
Horizontal flux $\Pi_\perp^{(\perp)}$ (resp.
$\Pi_\perp^{(\parallel)}$), transport $A_\perp^{(\perp)}$ (resp.
$A_\perp^{(\parallel)}$), diffusion $D_\perp^{(\perp)}$ (resp.
$D_\perp^{(\parallel)}$) and dissipation rate
$\tilde{\epsilon}_\perp^{(\perp)}$ (resp.
$\tilde{\epsilon}_\perp^{(\parallel)}$) of horizontal energy
(resp. vertical energy). $S^{(\perp)} =-\Pi_\perp^{(\perp)}
+D_\perp^{(\perp)} +A_\perp^{(\perp)}$ and $S^{(\parallel)}
=-\Pi_\perp^{(\parallel)} +D_\perp^{(\parallel)}
+A_\perp^{(\parallel)}$ are the sums of the measurable right-hand
side terms in the energy budgets~(\ref{eq:khm2d}) and
(\ref{eq:Yaglom}) respectively.} In (a), the horizontal flux of
horizontal energy is inverse, from small to large scales
($\Pi_\perp^{(\perp)} > 0$), whereas in (b) the horizontal flux of
vertical energy is direct, from large to small scales
($\Pi_\perp^{(\parallel)} < 0$).} \label{fig:perp}
\end{figure}

\subsubsection{Horizontal kinetic energy budget}

Let us first consider the energy budget for the
horizontal kinetic energy. In statistically-steady state,
Eq.~(\ref{eq:KHM_inhom_moyX_2D3C}) writes
\begin{equation}
\langle \overline{\tilde{\epsilon}_\perp^{(\perp)}}\rangle = - \Pi_\perp^{(\perp)} +
D_\perp^{(\perp)} + A_\perp^{(\perp)} + W_p + NT^{(\perp)}, \label{eq:khm2d}
\end{equation}
with
\begin{equation}
\langle \overline{\tilde{\epsilon}_\perp^{(\perp)}}\rangle = \frac{\nu}{2}
\langle \overline{(\partial_\alpha u_\beta + \partial_\beta u_\alpha)^2}\rangle, \qquad D_\perp^{(\perp)} =
\frac{\nu}{2}{\boldsymbol \nabla}_{\perp}^2 \langle
\overline{(\delta{\bf u}_{\perp})^2}\rangle,
\end{equation}
\begin{equation}
A_\perp^{(\perp)}=-\frac{1}{4V_X} \oint_{S_X}\overline{(\delta {\bf u}_{\perp})^2\tilde{\bf u}_{\perp}}\cdot d{\bf S_X},  \qquad W_p=-\frac{1}{2V_X} \oint_{S_X}\overline{\delta p\delta{\bf u}_{\perp}} \cdot d {\bf S}_X,\\
\label{eq:KHM_inhom_moyX_approx}
\end{equation}

The horizontal flux of horizontal energy $\Pi_\perp^{(\perp)}$ is
defined in Eq.~(\ref{eq:decompPi}), and $\epsilon_\perp^{(\perp)}$
is the dissipation of the horizontal velocity by the horizontal
shearing (with summation over  $\alpha, \beta = x,y$), such that
$D_\perp^{(\perp)} (r_\perp \rightarrow 0) =
\epsilon_\perp^{(\perp)}$. In Eq.~(\ref{eq:khm2d}), $NT^{(\perp)}$
contains the  viscous contribution from inhomogeneities, which we
neglect in the following (it is at least 2 orders of magnitude
smaller than the other terms), and the vertical transport, which
we cannot compute from 2D3C measurements. Here $\langle\,.\,
\rangle$ still denotes the spatial average over ${\bf X} \in V_X$.
Under the assumption of weak inhomogeneity, the velocity-pressure
flux $W_p$ is expected to be small compared to the transport
$A_\perp^{(\perp)}$ (see Appendix~\ref{sec:app}), and the
remaining terms in Eq.~(\ref{eq:khm2d}) can be readily computed
from the 2D-3C PIV measurements.

Figure~\ref{fig:perp}(a) shows the three measurable terms of the
r.h.s. of Eq.~(\ref{eq:khm2d}) together with their sum,
$S^{(\perp)}=-\Pi_\perp^{(\perp)}+D_\perp^{(\perp)}+A_\perp^{(\perp)}$.
These terms are averaged over the azimuthal angle $\varphi$. We
observe a good agreement between
$\langle\overline{\tilde{\epsilon}_{\perp}^{(\perp)}}\rangle$ and
the sum $S^{(\perp)}$ for scales smaller than $40$~mm, to within
20\%. The approximation of locally homogeneous and axisymmetric
turbulence, with a negligible velocity-pressure flux $W_p$, thus
seems valid at small scales. For larger scales we observe a
significant departure between
$\langle\overline{\tilde{\epsilon}_{\perp}}\rangle$ and
$S^{(\perp)}$, up to a factor of 2 at $r_\perp \simeq 60$~mm: for
such large scales the turbulent flow cannot be considered as
locally homogeneous anymore: the non-measured  pressure term $W_p$
cannot be neglected, so the interpretation of
$\Pi_\perp^{(\perp)}$ as a scale-by-scale energy transfer becomes
incorrect.

Focusing on scales smaller than $40$~mm, the advection of
horizontal kinetic energy $A_\perp^{(\perp)}$ is the only source
term in the energy budget. This source term is maximum at large
scales, as expected, but it remains significant over the whole
range of scales, suggesting a broad-band energy injection in this
system. If $E(r_\perp)$ is interpreted as a cumulative energy for
eddies of horizontal scale $r_\perp$ or less, the scale-by-scale
energy density has the form $dE / dr_\perp$, so the corresponding
forcing density is $dA_\perp / dr_\perp$. Since we observe
$A_\perp \simeq r_\perp^{1.5}$ at intermediate scales, the forcing
density scales as $r_\perp^{0.5}$, and it remains significant over
the range of scales considered here. This broad-band energy
injection is an important feature of boundary-forced inhomogeneous
turbulence, and is in strong contrast with the narrow-band forcing
often considered in numerical simulations of homogeneous rotating
turbulence.

Although the inverse cascade is evident from this horizontal
energy budget, it must be noted that its magnitude remains
moderate: $\Pi_\perp^{(\perp)}$ is never the dominant contribution
to the budget, even at the crossover between the viscous diffusion
and the forcing. A well developed inverse cascade over a wide
range of quasi-homogeneous scales at much larger Reynolds number
would be characterized by $\Pi_\perp^{(\perp)} \simeq A_\perp \gg
\langle\overline{\tilde{\epsilon}_{\perp}^{(\perp)}}\rangle$.
Here, the amount of energy transferred to large scales remains at
best of the same order as the small-scale viscous dissipation
$\langle\overline{\tilde{\epsilon}_{\perp}^{(\perp)}}\rangle$.

\subsubsection{Vertical kinetic energy budget}

We now consider the vertical kinetic energy budget using Eq.~(\ref{eq:Yaglom_inhom_2D3C}) in statistically-steady state:
\begin{equation}
\langle \overline{\tilde{\epsilon}_\perp^{(\parallel)}}\rangle = - \Pi_\perp^{(\parallel)} +
D_\perp^{(\parallel)} + A_\perp^{(\parallel)} + NT^{(\parallel)}, \label{eq:Yaglom}
\end{equation}
where
\begin{equation}
\langle \overline{\tilde{\epsilon}_\perp^{(\parallel)}}\rangle =
\nu \langle \overline{(\nabla_\perp u_\parallel)^2}\rangle, \qquad D_\perp^{(\parallel)} =
\frac{\nu}{2}{\boldsymbol \nabla}_{\perp}^2 \langle
\overline{(\delta{ u}_{\parallel})^2}\rangle,
\end{equation}
\begin{equation}
A_\perp^{(\parallel)}=-\frac{1}{4V_X} \oint_{S_X}\overline{(\delta { u}_{\parallel})^2\tilde{\bf u}_{\perp}}\cdot d{\bf S_X} \, ,
\label{eq:Yaglom_inhom_moyX_approx}
\end{equation}
The horizontal flux of vertical energy $\Pi_\perp^{(\parallel)}$
is defined in Eq.~(\ref{eq:decompPi}), and
$\epsilon_\perp^{(\parallel)}$ is the dissipation of the vertical
velocity by the horizontal shearing, with $D_\perp^{(\parallel)}
(r_\perp \rightarrow 0) = \epsilon_\perp^{(\parallel)}$. The
viscous contribution from inhomogeneities $NT^{(\parallel)}$ is
once again discarded for simplicity. A key feature of
Eq.~(\ref{eq:Yaglom}) is that it does not involve pressure: all
the terms can therefore be readily measured from stereo-PIV
measurements and the equation should be satisfied for scales at
which quasi-two-dimensionality holds.

The terms of Eq.~(\ref{eq:Yaglom}) are shown in
Fig.~\ref{fig:perp}(b), together with the sum of the right-hand
side terms
$S^{(\parallel)}=-\Pi_\perp^{(\parallel)}+D_\perp^{(\parallel)}+A_\perp^{(\parallel)}$.
Once again, there is a good overall agreement between $\langle
\overline{\tilde{\epsilon}_\perp^{(\parallel)}}\rangle$ and
$S^{(\parallel)}$ for $r_\perp \leq 40$~mm, clearly indicating a
direct cascade of vertical kinetic energy. The picture here is
simpler than for the horizontal energy budget: the vertical energy
is advected into the control domain ($A_\perp^{(\parallel)}$), it
is transferred by the nonlinearities to smaller scales
($\Pi_\perp^{(\parallel)}$), and it is dissipated at small scales
by viscosity ($D_\perp^{(\parallel)}$). We can assume that the
instabilities of the vortex dipoles generated near the flaps and
away from the measurement domain is the source of vertical energy
at large scale. Note that for very large Reynolds numbers, an
inertial range of scales $r_\perp$ with constant flux of vertical
energy is expected, characterized by
$\Pi_\perp^{(\parallel)}(r_\perp) = -
\epsilon_\perp^{(\parallel)}$. Although this is not achieved at
the moderate Reynolds number of this experiment, we can
nonetheless define a significant range of scales, centered around
10-20~mm, for which $-\Pi_\perp^{(\parallel)}$ is close to
$\epsilon_\perp^{(\parallel)}$ to within 20\%.

\subsubsection{Large-scale energy dissipation}

For the largest rotation rate, the scenario of a double cascade,
towards small scales for the vertical energy and towards large
scales for the horizontal energy, is well established from
Fig.~\ref{fig:perp}(a,b), at least for scales $r_\perp \leq
40$~mm. At larger scales, the departure between $\langle
\overline{\tilde{\epsilon}_\perp^{(\perp)}}\rangle$ and
$S^{(\perp)}$ may originate both from a departure from
two-dimensionality or from the non-measured velocity-pressure
correlation. Although less pronounced, there is also a discrepancy
between $\langle
\overline{\tilde{\epsilon}_\perp^{(\parallel)}}\rangle$ and
$S^{(\parallel)}$ at large scale, which indicates a departure from
a pure 2D3C state and possibly an influence of the horizontal top
and bottom boundaries.

We are therefore left with the following question: what sink of
energy absorbs the inverse horizontal energy flux at large scales?
A first candidate is Ekman friction on the horizontal boundaries.
Assuming that the flow is 2D3C in the central region of the tank,
we consider the laminar scaling for the Ekman layer thickness,
$\delta_E \simeq \sqrt{\nu/\Omega}$, and we deduce a typical
energy dissipation $\epsilon_{\Omega} = \sqrt{\nu \Omega}
u_\perp^2 / H$, where $H$ is the water depth and $u_\perp$ the
characteristic horizontal velocity. For $\Omega=16$~rpm, we obtain
$\epsilon_{\Omega} \simeq 10^{-7}$~m$^2$~s$^{-3}$, which turns out
to be of the order of the other terms of Eq.~(\ref{eq:khm2d}) (see
Fig.~\ref{fig:perp}). A significant fraction of the input
horizontal kinetic energy could therefore be dissipated at large
scale through Ekman friction. However, this order of magnitude
strongly relies on the boundary layers being laminar, which seems
questionable in the present experiment.

As an alternate explanation for the energy sink at large scales,
we may invoke a feedback of the large-scale flow on the forcing
device: the large-scale flow resulting from the inverse cascade
induces large-scale pressure forces that transfer some kinetic
energy back to the flaps, reducing the overall energy input in the
system. In this scenario, the flaps inject energy at intermediate
scales and receive energy from the large-scale flow through the
work of the pressure forces. In the framework of
Eq.~(\ref{eq:khm2d}), the corresponding sink of energy is taken
into account by the spatial flux and pressure terms: energy at
large scales is transferred outside of the control volume, towards
the flaps. Unfortunately, testing this scenario would require a
precise measurement of the energy input by the flaps and is beyond
the scope of the present study.

\section{Conclusion}

In this paper we provide experimental evidence of a double energy
cascade, direct at small scales and inverse at large scales, in a
forced rotating turbulence experiment. Since turbulence is
statistically steady, the inverse cascade does not manifest
through a temporal growth of kinetic energy, but it is
characterized by a change of sign of the scale-by-scale energy
flux. As the rotation rate is increased, the inverse cascade
becomes more pronounced and spreads down to the smallest scales.
As compared to previous experimental observations of an inverse
cascade in rotating
turbulence,\cite{Baroud2002,Morize2005,Yarom2013} here we provide
for the first time a direct scale-by-scale measurement of the
energy transfers in the horizontal plane. This allows us to
distinguish between the horizontal transfers of vertical and
horizontal kinetic energy. At the largest rotation rate, this
double cascade of the total energy can be described as the
superposition of an inverse cascade of horizontal energy and a
direct cascade of  vertical energy. This is consistent with the
2D3C dynamics expected in the limit of small Rossby numbers, with
the vertical velocity behaving as a passive scalar transported by
the horizontal flow.

Contrary to numerical simulations, in which energy is usually
supplied by a homogeneous body force acting on a prescribed narrow
range of wave numbers, in most experiments and in many natural
flows, energy is supplied at the boundaries. For a control domain
away from those boundaries, energy is advected from the boundaries
into the domain. This spatial flux of energy, which is strongly
related to the inhomogeneities of the turbulent statistics,
results in an effective broad-band energy injection term. In order
to interpret the energy transfers in such an experiment, it is
therefore necessary to separate the contributions from the spatial
transport and from the scale-by-scale transfers. We have performed
this analysis using the inhomogeneous generalization of the KHM
equation proposed by Hill,\cite{Hill2002} and we have measured
directly the energy transport term for scales at which the
velocity-pressure correlations can be neglected (quasi-homogeneous
approximation). Because of this effective broad-band forcing, the
inversion scale, which separates the direct and inverse cascades,
is not directly prescribed by the geometry of the forcing device
and decreases with the imposed rotation rate. Modelling this
inversion scale as a function of the Rossby number and forcing
geometry remains an open issue of first interest for flows of
oceanic and atmospheric relevance, such as convectively-driven
rotating flows.

\acknowledgments

We acknowledge P. Augier, P. Billant and J.-M. Chomaz for kindly
providing the flap apparatus, and A. Aubertin, L. Auffray, C.
Borget and R. Pidoux for their experimental help. This work is
supported by the ANR grant no. 2011-BS04-006-01 ``ONLITUR''. The
rotating platform ``Gyroflow'' was funded by the ``Triangle de la
Physique''.

\appendix

\section{Locally homogeneous and axisymmetric turbulence}
\label{sec:app}

In this appendix we show how one can neglect the velocity-pressure
correlations in the inhomogeneous KHM equations
(\ref{eq:KHM_inhom_moyX})-(\ref{eq:KHM_inhom_moyX_2D3C}) under the
assumptions of local homogeneity and axisymmetry.

Let us first consider that the turbulence is invariant to a
reflection with respect to a horizontal plane (non-helical
turbulence). Let us then denote as $L_\text{inh}$ the typical
scale of the inhomogeneity of the turbulence statistics and focus
on small separations $r \ll L_\text{inh}$, or equivalently $r /
L_\text{inh} = \zeta \ll 1$. We decompose the velocity and
pressure fields into ${\bf u}={\bf V}+{\bf v}$ and $p=P+q$, where
${\bf V}$ and $P$ contain the large scales of the flow and ${\bf
v}$ and $q$ are the small-scale fluctuations. One can think about
${\bf V}$ and $P$ as coarse-grained versions of the velocity and
pressure fields on a scale that is smaller than $L_\text{inh}$ but
larger than the range of scales $r$ of interest. Then ${\bf
v}={\bf u}-{\bf V}$ and $q=p-P$ are the small-scale remainders.

Local axisymmetry and homogeneity consists in assuming that the
statistics of ${\bf v}$ and $q$ are axisymmetric and homogeneous
for separations $r \ll L_\text{inh}$. Let us evaluate the
different terms in the integrand of (\ref{eq:phiinh}) under this
assumption.

The pressure term decomposes into
\begin{equation}
\left<\delta {\bf u} \delta p \right> = \left<\delta {\bf V} \delta P \right>+\left<\delta {\bf V} \delta q \right>+\left<\delta {\bf v} \delta P \right>+\left<\delta {\bf v} \delta q \right> \, .
\end{equation}
Because $P$ and ${\bf V}$ evolve on a spatial scale $r \ll L_\text{inh}$,
\begin{equation}
\delta P \simeq {\bf r} \cdot {\boldsymbol \nabla} P \sim \frac{r}{L_\text{inh}} P= \zeta P \, .
\end{equation}
Similarly, $\delta {\bf V} \sim \zeta {\bf V}$. Hence
$\left<\delta {\bf V} \delta P \right>= \mathcal{O}(\zeta^2)$,
$\left<\delta {\bf V} \delta q \right>= \mathcal{O}(\zeta)$ and
$\left<\delta {\bf v} \delta P \right>= \mathcal{O}(\zeta)$. We
deal with the term $\left<\delta {\bf v} \delta q \right>$ using
the assumption of local homogeneity and axisymmetry: under a
rotation of angle $\pi$ around a vertical axis passing through the
mid-point ${\bf X}$, followed by a reflection with respect to the
horizontal plane containing ${\bf X}$, ${\bf v}$ becomes ${\bf
v}'$, ${\bf v}'$ becomes $-{\bf v}$, $q'$ becomes $q$ and $q$
becomes $q'$. Hence $ \delta {\bf v} \delta q $ becomes $- \delta
{\bf v} \delta q$, so that on statistical average this quantity
vanishes: $\left<\delta {\bf v} \delta q \right>=0$. As a
consequence, the velocity-pressure correlation term $\left<\delta
{\bf u} \delta p \right>$ is of order $\mathcal{O}(\zeta)$ in the
weakly inhomogeneous limit. We therefore expect $\left<\delta {\bf
u} \delta p \right>$ to be negligible compared to the transport
term $\left< (\delta {\bf u})^2 \tilde{\bf u} \right>$, which is
of order $\mathcal{O}(\zeta^0)$.

The viscous term in the integrand of (\ref{eq:phiinh}) is of order
\begin{equation}
\nu \frac{\left<(\delta {\bf u})^2 \right> }{ L_\text{inh}} \sim \frac{\nu}{|\tilde{\bf u}| L_\text{inh}} \left< (\delta {\bf u})^2 \tilde{\bf u} \right>\, .
\end{equation}
It is therefore negligible compared to the advective term of the
integrand, because the Reynolds number based on $L_\text{inh}$ is
large. Note that because $r \ll L_\text{inh}$, derivatives with
respect to $r$ are much larger than derivatives with respect to
$X$. In the ensemble average of Eq.~(\ref{eq:KHM_inhom_nomoy}),
the ``inhomogeneous" viscous contribution can be estimated as $
\nu {\boldsymbol \nabla}_{{\bf X}}^2 \left<(\delta {\bf u})^2
\right> \sim \zeta^2  \nu {\boldsymbol \nabla}_{{\bf r}}^2
\left<(\delta {\bf u})^2 \right> $, hence it is negligible
compared to the ``homogeneous" viscous contribution $2 \nu
{\boldsymbol \nabla}_{{\bf r}}^2 \left<(\delta {\bf u})^2 \right>
$. For weakly inhomogeneous turbulence, one can therefore keep the
latter while neglecting the former.

To conclude, the advective term $\left< (\delta {\bf u})^2
\tilde{\bf u} \right>$ is the dominant term in the integrand of
Eq.~(\ref{eq:phiinh}). For weakly inhomogeneous turbulence, one
can therefore retain only this term of the integrand, which
ensures the injection of kinetic energy into the control volume.

\end{document}